\documentclass[aps,pra,showpacs,twocolumn,groupedaddress]{revtex4}
\usepackage{graphicx}
\usepackage{CJK}
\usepackage{amsmath}
\usepackage{amssymb}

\begin{document}
\begin{CJK*}{GBK}{song}

\title{Time-dependent self-trapping of Bose-Einstein
Condensates in a double-well potential}
\author{B. Cui, L. C. Wang, X. X. Yi\footnote{Email:yixx@dlut.edu.cn}}
\affiliation{School of Physics and Optoelectronic Technology,\\
Dalian University of Technology, Dalian 116024 China}

\date{\today}

\begin{abstract}
Based on  the mean-field approximation and  the phase space
analysis, we discuss  the dynamics of Bose-Einstein condensates in a
double-well potential. By applying a periodic modulation to the
coupling between the condensates, we find the condensates can be
trapped in the time-dependent eigenstates of the effective
Hamiltonian, we refer to this effect as time-dependent self-trapping
of BECs. A comparison of this self-trapping with the adiabatic
evolution is made, finding that the adiabatic evolution beyond the
traditional(linear) adiabatic condition can be achieved in BECs by
manipulating the nonlinearity and the ratio of the level bias to the
coupling constant. The fixed points for the system are calculated
and discussed.
\end{abstract}

 \maketitle

\section{introduction}

Bose-Einstein condensates (BECs) in a double-well potential have
attracted much attention in the past decades, it provides a useful
tool to study fundamental problems in quantum physics at the
macroscopic scale and opens the possibility of practical
applications, such as high-precision measurements, interferometry
and thermometry \cite{andrews97}. Different from Josephson junctions
realized in superconductors or superfluids
\cite{Pereverzev97,sukhatme01}, the interatomic interactions in BECs
play an important role in its dynamics, leading to many rich and
interesting nonlinear effects. Self-trapping
\cite{smerzi97,milburn97,raghavan97,albiez05} is one of these
interesting phenomena from  which much attention has been received
in recent years. When the initial population imbalance between the
two wells and the nonlinearity in BECs is above a critical value,
the amplitude of the Josephson oscillations are extremely compressed
and most of the atoms are trapped in one well as theoretically
predicted and experimentally observed \cite{albiez05}.

The self-trapping is originally defined and studied for a
time-independent system (Hamiltonian). Consider a system governed by
a time-dependent Hamiltonian, the following questions naturally
arise: How can we make the self-trapping happen at a time-dependent
state? What are the fixed points in this situation? How do these
fixed points behave? We will answer these questions in this paper.
In fact, adiabatic dynamics in BECs has been investigated in
different regimes, such as Landau-Zener transition
\cite{wu00,liu02,liu03,theocharis06,graefe06,wang06,
itin07,altland09,smith09}, Rosen-Zener process \cite{ye08,fu09}.
Meanwhile, several proposals  have been proposed with the stimulated
Raman adiabatic passage technique(STIRAP) \cite{bergmann98}, to
coherently manipulate BECs in double-well or triple-well potentials
by adiabatically following a spatial dark state of the system
\cite{eckert04,vitanov06,rab08,nesterenko09,ottaviani10}, finding
that nonlinearity plays a negative role against adiabatic evolution
in some regimes \cite{ottaviani10}. Different from these works, we
discuss here the possibility of manipulating the BECs to
adiabatically follow the instantaneous  eigenstate of the system
with the help of atom-atom couplings.  We find that nonlinearity can
help the system to  track the instantaneous eigenstates of the
system,  leading  to a self-trapping. By manipulating the
nonlinearity and the ratio of  the level bias to the coupling
constant,   adiabatic evolution beyond the traditional adiabatic
condition can be achieved in BECs in both symmetric and asymmetric
double-well potentials.

The paper is organized as follows. In Sec. \ref{fml}, we introduce
the model and transform the equations to the time-dependent
representation, choosing the instantaneous eigenstates as the basis.
In Sec. \ref{sy} and Sec. \ref{asy}, we investigate the
self-trapping in BECs in a symmetric and an asymmetric double-well
potential respectively,  and discuss the effect of nonlinearity on
the dynamics. Finally, we conclude our results in Sec. \ref{con}.

\section{model}\label{fml}

We start with  the  standard nonlinear two-level model, which
describes a BEC in a double-well potential \cite{wu00,liu02,wang06}.
In a time-independent basis $\{|L\rangle, |R\rangle\}$, the model
can be written as \cite{smerzi97,milburn97}
\begin{eqnarray}
i\frac{d}{dt}\begin{pmatrix}a\\b\end{pmatrix}&&=\frac{1}{2}
\begin{pmatrix}r+ms&v\\v&-r-ms\end{pmatrix}\begin{pmatrix}a\\b\end{pmatrix}\nonumber\\
&&\equiv H_0\begin{pmatrix}a\\b\end{pmatrix},\label{model}
\end{eqnarray}
where $s=|b|^2-|a|^2$ stands  for the population imbalance between
the two wells denoted respectively by $|L\rangle$ and $|R\rangle.$
$r$, $v$ and $m$ characterize the level bias, the tunneling between
the two wells and the nonlinearity in BECs, respectively.  Here and
hereafter, we rescale $m$, $r$, $v$ in units of $\omega$,  and $t$
in units of $1/\omega$, $\hbar=1$ has been set, hence all parameters
in this paper are of dimensionless. In previous study, the tunneling
$v$ is a real parameter, here we discuss a complex tunneling
dependent on time through $V=ve^{i\omega{t}}$, which can be viewed
as a nonlinear two-level system  driven by a time-dependent
magnetic field \cite{zhang08}, $\vec{B}=(B_x,B_y,B_z).$ There are
many ways to realize experimentally such a Hamiltonian. For
instance, one may simulate this Hamiltonian by a BEC in a
double-well potential with the height of the potential barrier
modulated and the phases of the two modes mismatched.
 Define $
\cos{\beta}=\frac{r}{\sqrt{r^2+v^2}}, \alpha=2B_0 =\sqrt{r^2+v^2},
\phi=\omega{t},$ the effective Hamiltonian $H_0$ \textbf{(replace $v$ by
$V=ve^{i\omega{t}}$ and $V^{*}$ respectively in Eq.(\ref{model}))} can be rewritten as,
\begin{eqnarray}
H_0 &=&
B_0(\sin\beta\cos\phi\sigma_x+\sin\beta\sin\phi\sigma_y\nonumber\\
&+&\cos\beta\sigma_z)+\frac 1 2 ms \sigma_z,
\end{eqnarray}
where $B_0$ acts as  the amplitude of the magnetic  field,  $\sigma_x$,
$\sigma_y$ and $\sigma_z$ are the Pauli matrices for the two-level
system. $\beta$ behaves as  the
ratio of the level bias to the coupling constant, $\alpha$
characterizes the amplitude of the driving field, and $\phi$
represents the swept angle of driving field.

It is easy to write the instantaneous  eigenstates of the system,
$|E_{+}\rangle=\begin{pmatrix}\cos{\frac{\beta}{2}}
,&\sin{\frac{\beta}{2}}e^{i\omega{t}}\end{pmatrix}^{T}$ and
$|E_{-}\rangle=\begin{pmatrix}-\sin{\frac{\beta}{2}}e^{-i\omega{t}}
,&\cos{\frac{\beta}{2}}\end{pmatrix}^{T}$ in the basis $|L\rangle$
and $|R\rangle$, with the corresponding eigenenergy
$E_{\pm}=\pm\alpha/2$. After  a representation transformation by
replacing the original time-independent  biases $|L\rangle$ and
$|R\rangle$ with the instantaneous energy eigenstates
$|E_{+}\rangle$ and $|E_{-}\rangle$, we obtain a new set of equation
\begin{equation}
i\frac{d}{dt}\begin{pmatrix}c\\d\end{pmatrix}=\begin{pmatrix}E_{+}+M_1{s}+E_{++}
&{M_2}{s}+E_{+-}\\{M_2}^{\ast}{s}+E_{-+}&E_{-}-M_1{s}+E_{--}\end{pmatrix}
\begin{pmatrix}c\\d\end{pmatrix} \label{dotcd}
\end{equation}
with $c$ and $d$ representing the amplitude of the BECs in
$|E_{+}\rangle$ and $|E_{-}\rangle$, respectively, i.e.,
$|\psi\rangle=c| E_{+}\rangle+d|E_{-}\rangle.$ The following
notations have been used,
\begin{eqnarray}
s&=&\cos{\beta}s_1+\sin{\beta}s_2\nonumber\\
&\equiv&\cos{\beta}(|d|^2-|c|^2)+\sin{\beta}(cd^{\ast}e^{i\phi}
+c^{\ast}de^{-i\phi}),\nonumber
\end{eqnarray}
\begin{eqnarray}
M_1&=&\frac{m}{2}\cos{\beta},\nonumber\\
M_2&=&-\frac{m}{2}\sin{\beta}e^{-i\phi},\nonumber\\
E_{++}&=&-E_{--}\nonumber\\
&=&-i\langle
E_+|\dot{E}_{+}\rangle\nonumber\\
&=&\frac{\omega}{2}(1-\cos{\beta}),\nonumber\\
E_{+-}&=&E_{-+}^{\ast}\nonumber\\
&=&-i\langle
E_+|\dot{E}_{-}\rangle\nonumber\\
&=&\frac{\omega}{2}\sin{\beta}e^{-i\phi}.
\end{eqnarray}
Here $s$ representing the nonlinearity consists of  two types,
namely $s_1$ and $s_2$, which correspond to population imbalance and
spatial atomic coherence \cite{ottaviani10}, respectively. $M_1$ and
$M_2$ characterize the strength of nonlinearity in the system.
$E_{\pm}$ and $E_{\pm\pm}$ are actually the integrand  of the
dynamical phase and Berry phase \cite{simon83,berry84}. $E_{+-}$ and
$E_{-+}$ are terms that may induce  tunneling between the new bases
$|E_{+}\rangle$ and $|E_{-}\rangle$, and they are also the parts
which lead the adiabatic theorem to break down \cite{born28} for the
system without nonlinearity. Observing Eq.(\ref{dotcd}), we find
that the nonlinearity may help to track the system on the
instantaneous eigenstate of the Hamiltonian, meanwhile the change
rate of the Hamiltonian can affect the self-trapping of the BECs on
the instantaneous eigenstates, how do they relate to each other?

With the mean-field approximation, the probability amplitudes can be
written as $c=|c|e^{i\theta_c}$ and $d=|d|e^{i\theta_d}$. By
defining population imbalance and relative phase as $Z=|d|^2-|c|^2$
and $\Theta=\theta_d-\theta_c$, respectively, we can study the
dynamics of the system in its (classical) phase space
\cite{smerzi97,liu03}. However, the situation under study is little
bit different, the BECs are driven effectively by a rotating
magnetic field with frequency $\omega$, this motivates us to define
{\it pseudo fixed points} by
\begin{eqnarray}
\dot{Z}=0,  \ \ \mbox{and}\ \ \dot{\Theta}=\omega,
\end{eqnarray}
in the phase space, which corresponds to  states of the form,
$|\psi\rangle=|c||E_+\rangle+|d|e^{i\omega{t}}|E_-\rangle$ with
fixed population difference in the language of wavefunction. We can
track this state  both in adiabatic and diabatic evolution. For
simplicity, we will refer the {\it pseudo fixed points} as fixed
points when no confusion arises. Define $\sin{\gamma}=|c|$ and
$\cos{\gamma}=|d|$, the state corresponding to the fixed point can
be represented in the basis $\{|L\rangle,|R\rangle\}$ as
$|\psi\rangle=\begin{pmatrix}\sin{(\gamma-\frac{\beta}{2})},&
\cos{(\gamma-\frac{\beta}{2})}e^{i\omega{t}}\end{pmatrix}^{T}$.

\begin{figure}
\includegraphics*[width=0.7\columnwidth,height=0.5\columnwidth]{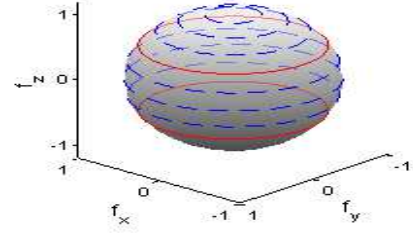}
\caption{(color online)  Illustration of self-trapping states on the
Bloch sphere. The Bloch vectors for these states are
$f_x=\sin{(2\gamma-\beta)}\cos{\phi}$,
$f_y=\sin{(2\gamma-\beta)}\sin{\phi}$ and
$f_z=\cos{(2\gamma-\beta)}$. The basis is $\{ |L\rangle, |R\rangle
\}.$  The red-solid lines denote the instantaneous  eigenstates of
the system and blue-dashed lines represent the other self-trapping
states. The parameters chosen are $r=1$, $v=\sqrt{3}$ and
$\beta=\pi/3$.}\label{blochsphere}
\end{figure}
\begin{figure*}
\includegraphics*[width=18cm]{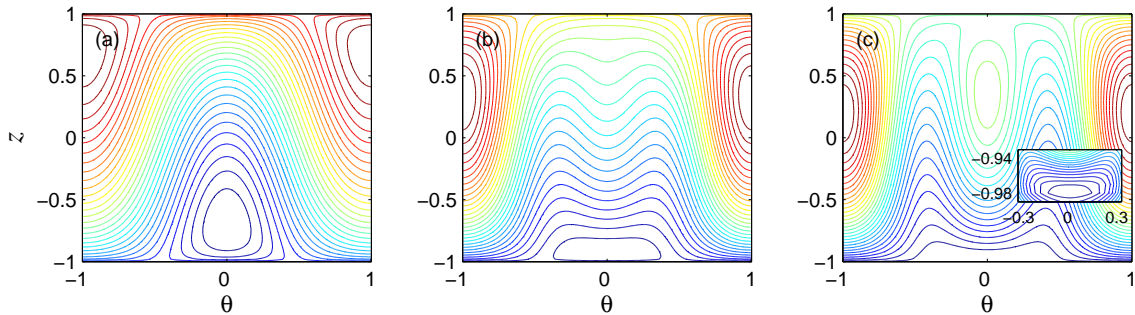}
\caption{Phase-space trajectories with different nonlinearity.
$\theta$ is in units of $\pi$.  $v=1$ and $\omega=1$. The
nonlinearity  is set as $m=0$ in (a), $m=2$ in (b) and $m=4$ in (c),
respectively. For weak nonlinearity, there exist two fixed points.
When $|M_2|>|E_{+-}|$ (strong nonlinearity), the other two fixed
points appear  in the phase-space. The system can be trapped in the
pole fixed point (we denote by the pole fixed point the point with
$Z$ very close to $\pm1$, i.e., near the pole on the Bloch sphere)
when nonlinearity is strong enough. The inset in (c) is the enlarged
circles around pole fixed point at the bottom of the phase
space.}\label{1}
\end{figure*}
By manipulating $\gamma$ (location of the fixed points), we can
obtain different instantaneous self-trapping states, which maintains
the population imbalance  and keeps the relative phase matched with
the driving field, as shown in Fig. \ref{blochsphere}. For example,
the fixed point $\gamma=0$ ($Z=1$) and $\gamma=\frac{\pi}{2}$
($Z=-1$) represent the instantaneous energy eigenstates
$|\psi\rangle=|E_{-}\rangle$ and $|\psi\rangle=|E_{+}\rangle,$
respectively.

\section{symmetric double-well potential}\label{sy}

\begin{figure}
\includegraphics*[width=0.8\columnwidth,height=0.9\columnwidth]{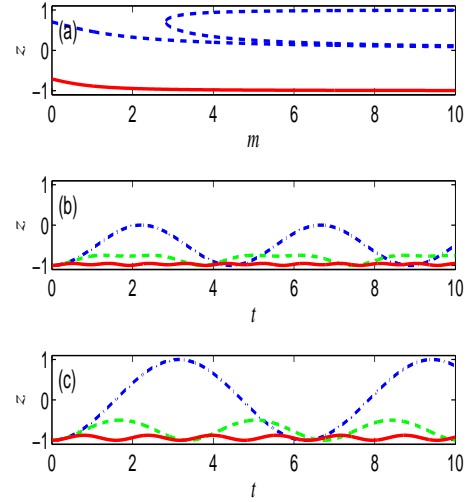}
\caption{(color online) (a) The distribution of the fixed points
with different nonlinearity. The red-solid line depicts the pole
fixed point  and blue-dashed lines denote the other fixed points.
Parameters chosen are $\omega=1$ and $v=1$. (b) Time evolution of
the population imbalance  with eigenstates of the Hamiltonian as the
initial state. Parameters chosen are $\omega=1$, $v=1.$  \
$m=0,2,10$ correspond to blue-dash-dotted line, green-dashed line
and red-solid line, respectively. (c) is the same as  (b), but with
different parameters, i.e.,  $\omega=1$, $v=0.01$ and $m=0,100,400$.
The system is trapped approximately in one of the instantaneous
eigenstates even if  adiabatic condition is broken,  as the
red-solid lines in (b) and (c) show.}\label{3}
\end{figure}

In this section, we focus on the BECs trapped in a  symmetric
double-well potential (i.e., $r=0$, $\beta=\pi/2$) and discuss the
self-trapping  with different nonlinearity. In this case, the
parameters in Eq. (\ref{dotcd})  reduces  to
\begin{eqnarray}
s&=&s_2, \nonumber\\
E_{\pm}&=&\pm
v/2,\nonumber\\
M_1&=&0,\nonumber\\
M_2&=&-\frac{m}{2}e^{-i\phi},\nonumber\\
E_{++}&=&\omega/2,\nonumber\\
E_{+-}&=&\frac{\omega}{2}e^{-i\phi},\label{sym}
\end{eqnarray}
where only the nonlinearity $s_2$ from the spatial atomic coherence
takes place. In this case,  Eq. (\ref{dotcd}) can be written  as
\begin{eqnarray}
i\frac{d}{dt}\begin{pmatrix}c\\d\end{pmatrix}=\frac{1}{2}\begin{pmatrix}v+\omega
&(-ms+\omega)e^{-i\phi}\\(-ms+\omega)e^{i\phi}&
-v-\omega\end{pmatrix}\begin{pmatrix}c\\d\end{pmatrix}.
\end{eqnarray}
The system in this limit  is interesting because it provides a model
to show the nonlinear effect when the nonlinearity appears only in
the off-diagonal term. The self-trapping and the adiabatic evolution
alike in the sense that  both of them can conserve the population on
a state. The difference is that the (traditional) self-trapping
happens at a time-independent state, while the adiabatic evolution
conserves the population on the time-dependent eigenstates of the
Hamiltonian. We now show that the self-trapping can appear at a
time-dependent state. By the adiabatic theorem, if the system starts
in an eigenstate of the Hamiltonian at time $t=0$, it will evolve to
the corresponding eigenstate of the Hamiltonian at later time as
long as the Hamiltonian changes slowly enough. For the considered
system without nonlinearity ($m=0$), the adiabatic condition
\cite{born28} reads
\begin{equation}
A=\omega/2v \ll 1,
\end{equation}
where $A$ is the adiabatic condition that we will refer to in later
discussion. From Eq. (\ref{sym}), we find that the change rate
($E_{+-}$) of the Hamiltonian affects the number and behavior of
fixed points, now we study this effect by the phase space analysis.
Define,
\begin{eqnarray}
z&=&Z=|d|^2-|c|^2,\nonumber\\
\theta&=&\Theta-\phi=\theta_d-\theta_c-\phi,\label{defz}
\end{eqnarray}
where $z$ and $\theta$ satisfy
\begin{eqnarray}
\dot{z}=&-\omega\sqrt{1-z^2}\sin{\theta}+\frac{m}{2}(1-z^2)\sin{2\theta},\nonumber\\
\dot{\theta}=&\frac{\omega{z}}{\sqrt{1-z^2}}\cos{\theta}-mz\cos^2{\theta}+v.
\end{eqnarray}
Then we can cast the dynamical system into a classical Hamiltonian
\begin{equation}
H_e(z,\theta)=vz-\frac{m}{2}z^2\cos^{2}{\theta}
+\frac{m}{4}\cos{2\theta}-\omega\sqrt{1-z^2}\cos{\theta}.
\end{equation}
By considering $\dot{\theta}=0$ and $\dot{z}=0$, we obtain the
following equation,
\begin{equation}
z^4-2\frac{v}{m}z^3-(1-\frac{v^2}{m^2}
-\frac{\omega^2}{m^2})z^2+2\frac{v}{m}z-\frac{v^2}{m^2}=0.
\label{symz}
\end{equation}
To study  the stability of the fixed points, we introduce the
Jacobian matrix \cite{strogatz94,boyce97}
\begin{equation}
J=\left(
\begin{array}{ccc}
 {\partial{\dot{z}}}/{\partial{z}} &&
 {\partial{\dot{z}}}/{\partial{\theta}} \\
 {\partial{\dot{\theta}}}/{\partial{z}} &&
 {\partial{\dot{\theta}}}/{\partial{\theta}} \\
\end{array}
\right).
\label{jacobian}
\end{equation}
It is well known that the eigenvalues of the Jacobian matrix
characterize  the type of  fixed points and determine the stability
of them. Two imaginary eigenvalues indicate a stable elliptic fixed
point, while two real eigenvalues correspond to an unstable
hyperbolic fixed point. So the stable elliptic fixed points can be
singled out  by calculating eigenvalues
of the corresponding Jacobian matrix.\\
\begin{figure}
\includegraphics*[width=0.7\columnwidth,height=0.5\columnwidth]{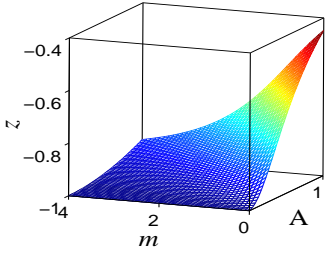}
\caption{The pole fixed point versus the nonlinearity $m$ and
$\omega$, when $v$ is fixed, $\omega$ is proportional to the
adiabatic condition $A$.}\label{2}
\end{figure}
From Eq. (\ref{symz}), we find that the number of the fixed points
depends on the ratio of the nonlinear parameter $M_2$ to the
adiabatic term $E_{+-}$. When nonlinearity term is smaller than the
adiabatic term, i.e., $|M_2|<|E_{+-}|$, there are only two fixed
points, while for strong nonlinearity ($|M_2|>|E_{+-}|$), the other
two  fixed points appear  in the phase space with proper coupling
constant $v$, this was  shown in Fig. \ref{1} and in Fig.
\ref{3}(a). This is interesting, as $E_{+-}$ in linear system can
induce population transfer between states $|E_+\rangle$ and
$|E_-\rangle$, while it might lead to bifurcation in nonlinear
systems. In what follows, we focus on the elliptic fixed point
nearest to $z=\pm1$. This fixed point is stable and describes
approximately  the instantaneous eigenstate of the system, hence we
call it pole fixed point. With strong nonlinearity $m$, the pole
fixed point is very close  to $z=\pm1$, see Fig. \ref{1} and Fig.
\ref{2}. In this sense we claim that nonlinearity favors the fixed
points $z=\pm 1$.

When $A\gg 1$, i.e., the adiabatic condition in linear case has been
completely broken down,   Eq. (\ref{symz}) can be solved
analytically up to  the zeroth-order in $v/m$ (we  assume  that
$m\gg{v}$), there exist three solutions given by
\begin{eqnarray}
z=\left\{\begin{aligned}
&\pm\sqrt{1-{\frac{{\omega}^2}{m^2}}}\\
&0
\end{aligned}\right..\label{largeAz}
\end{eqnarray}
Eq. (\ref{largeAz}) tells that we can manipulate the pole fixed
point by adjusting the nonlinearity $m$ to satisfy $m\gg{\omega}$ in
this limit, this is numerically confirmed  as shown in Fig.
\ref{3}(c).

\section{asymmetric double-well potential}\label{asy}

In this section, we discuss a more general case that the BECs are
trapped in an asymmetric double-well potential $(r\neq 0)$. With the
same definition for $z$ and $\theta$ in Eq. (\ref{defz}), we map the
system into a classical phase space. In this case, $z$ and $\theta$
satisfy
\begin{eqnarray}
\dot{z}&=&-\frac{\omega{v}}{\alpha}\sqrt{1-z^2}\sin{\theta}\nonumber\\
&+& \frac{mrv}{\alpha}z\sqrt{1-z^2}\sin{\theta}
+\frac{mv^2}{2\alpha^2}(1-z^2)\sin{2\theta},\nonumber\\
\dot{\theta}&=&\frac{\alpha^2-\omega{r}}{\alpha}
+\frac{mr^2}{\alpha^2}z+\frac{\omega{v}}
{\alpha}\frac{z}{\sqrt{1-z^2}}\cos{\theta}\nonumber\\
&+& \frac{mrv}{\alpha^2}\frac{1-2z^2}{\sqrt{1-z^2}}\cos{\theta}
-\frac{mv^2}{\alpha^2}z\cos^2{\theta}.
\end{eqnarray}
The effective Hamiltonian and the equation for $z$ can be analytically
expressed as
\begin{eqnarray}
H_e(z,\theta)&=&\frac{m}{2\alpha^2}z^2(r^2-v^2\cos^2{\theta})+\frac{mrv}
{\alpha^2}z\sqrt{1-z^2}\cos{\theta}\nonumber\\
&-&\frac{\omega{v}}{\alpha}\sqrt{1-z^2}\cos{\theta}+\frac{\alpha^2-\omega{r}}{\alpha}z\nonumber\\
&+&\frac{mv^2}{4\alpha^2}\cos{2\theta},
\end{eqnarray}
\begin{eqnarray}
&&m^2\alpha^4{z^4}+\alpha^4(\alpha^2+\omega^2-m^2-2\omega{r}){z^2}\nonumber\\
&&+2m\alpha^3(r^2-v^2-\omega{r}){z^3}+2m\alpha(\omega{r^3}-\alpha^2r^2+\alpha^2v^2){z}\nonumber\\
&&+m^2r^2v^2-\alpha^2(\alpha^2-\omega{r})^2=0.
\end{eqnarray}
\begin{figure}
\includegraphics*[width=1.0\columnwidth,height=1.0\columnwidth]{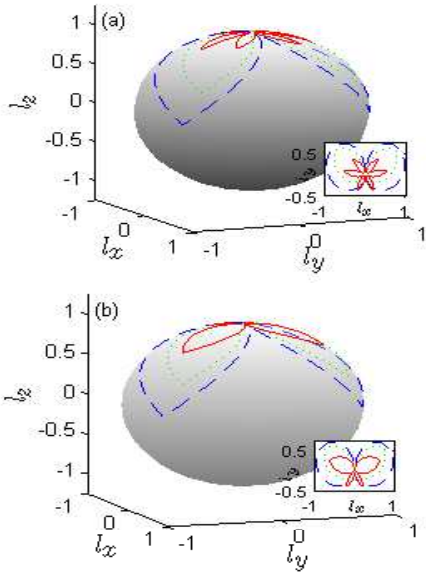}
\caption{(color online) (a) Illustration of the biggest-valued fixed
points (including stable and unstable ones) with different
nonlinearity.  (b) The  biggest-valued stable elliptic fixed points
(pole fixed point) with different nonlinearity. Here
$l_x=\sqrt{1-z^2}\cos{\beta}$, $l_y=\sqrt{1-z^2}\sin{\beta}$ and
$l_z=z$, where $z$ is the population imbalance. Parameters chosen
are $\omega=2$, $\alpha=1$ and $m=0,2,4$ corresponding to  blue-dashed,
green-dotted and red-solid lines.}\label{4}
\end{figure}
\begin{figure}
\includegraphics*[width=0.8\columnwidth,height=0.5\columnwidth]{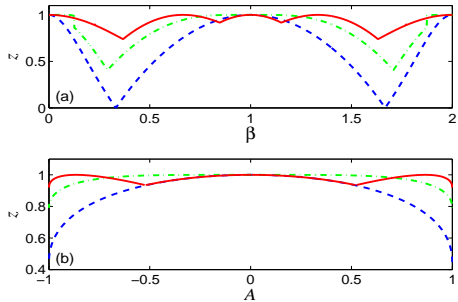}
\caption{(color online)  The pole fixed point versus $\beta$(see
(a)) and $A$ (the adiabatic condition parameter, see (b)). $\beta$
is in units of $\pi$. Parameters chosen are $\omega=2$, $\alpha=1$. $m=0,2,4$
are for the blue-dashed, green-dash-dotted and red-solid lines
respectively.}\label{5}
\end{figure}

By the phase space analysis, we find that, different from the
symmetric case, the fixed points are determined not only  by the
nonlinearity $m$ but also by the ratio of the level bias $r$ to the
coupling constant $v$. Without  nonlinearity, the adiabatic
condition \cite{born28} is
\begin{eqnarray}
A=\frac{\omega}{2\alpha}\sin{\beta}\ll1.
\end{eqnarray}

Now we investigate the dynamics of the system  under different
nonlinearity $m$ and  $\beta$, which have the same notations as
before, while $\omega$ and $\alpha$ are fixed corresponding to a
constant frequency and amplitude of the driving field. The adiabatic
condition becomes $A=\sin{\beta}\ll1$, when we set
$\omega/{2\alpha}=1$. By examining the location of the
biggest-valued fixed points with different nonlinearity from
adiabatic regime to diabatic regime, we find that the nonlinearity
in BEC does play a positive role in the process of adiabatic
evolution as shown in Fig. \ref{4} and Fig. \ref{5}. When
nonlinearity is weak, it pushes the pole fixed point to $z=\pm1$ at
all circumstances, as shown by the green-dotted line in Fig.
\ref{4}(b). When the nonlinearity becomes strong, bifurcation
appears, corresponding to the occurrence of new fixed points in the
phase space and the system can evolve on its instantaneous energy
eigenstate beyond adiabatic condition at some special  $\beta$, as
shown by the red-solid line in Fig. \ref{5}(a). Before closing this
section, we would like to address that all fixed points can be
easily calculated and analyzed. Aiming to study the self-trapping in
the instantaneous eigenstates of the Hamiltonian, we here mainly
focus on the fixed point nearest to $z=\pm 1$, discussions on the
other fixed points can be carried out in the same way.

Two remarks are now in order. (1) The linear and nonlinear systems
discussed here are not independent, the linear system is exactly a
limiting case of the nonlinear system with zero nonlinearity. (2)
The fixed points are defined based on  the basis spanned by the
instantaneous eigenstates, so the imbalance of the population
represented by $z$ denotes the population difference on the two
instantaneous eigenstates, it ranges from $-1$ to $1$ depending on
the system parameters. Moreover, the population on the right and
left wells can also be manipulated, this can be found via the
definition of the instantaneous eigenstates.

\section{conclusion}\label{con}

In summary, the time-dependent self-trapping for Bose-Einstein
condensates in a double-well potential has been  introduced and
studied in this paper. Both the atom-atom coupling and the quality
which characterize the change of the system play important roles in
the self-trapping. Fixed points are calculated and discussed. These
results suggest that the nonlinearity can help tracking the
instantaneous eigenstates of the time-dependent Hamiltonian,
providing a way to manipulate quantum systems.

We are grateful to Dr. Jiangbin Gong for helpful suggestions on
early versions of this manuscript. This work is supported by NSF of
China under grant Nos. 10775023 and 10935010.

\end{CJK*}

\begin{references}

\bibitem{andrews97} M. R. Andrews, C. G. Townsend, H. J. Miesner, D. S. Durfee,
D. M. Kurn, and W. Ketterle, Science {\bf 275}, 637 (1997); H. J.
Wang, X. X. Yi, and X. W. Ba, Phys. Rev. A {\bf 62}, 023601 (2000);
Y. Shin, M. Saba, A. Schirotzek, T. A. Pasquini, A. E. Leanhardt, D.
E. Pritchard, and W. Ketterle, Phys. Rev. Lett. {\bf 92}, 150401
(2004); J. Est\`{e}ve, J. B. Trebbia, T. Schumm, A. Aspect, C. I.
Westbrook, and  I. Bouchoule, Phys. Rev. Lett. {\bf 96}, 130403
(2006); G. F. Wang, L. B. Fu, and J. Liu, Phys. Rev. A {\bf 73},
013619 (2006); B. V. Hall, S. Whitlock, R. Anderson, P. Hannaford,
and A. I. Sidorov, Phys. Rev. Lett. {\bf 98}, 030402 (2007); U.
Hohenester, P. K. Rekdal, A. Borzi, and J. Schmiedmayer, Phys. Rev.
A {\bf 75}, 023602 (2007);  J. Esteve, C. Gross, A.Weller, S.
Giovanazzi, and M. K. Oberthaler, Nature (London) {\bf 455}, 1216
(2008).

\bibitem{Pereverzev97} S. V. Pereverzev, A. Loshak, S. Backhaus, J. C. Davis,
and R. E. Packard, Nature (London) {\bf 388}, 449 (1997).

\bibitem{sukhatme01} A. K. Sukhatme, Y. Mukharsky,
T. Chui, and D. Pearson, Phys. Rev. Lett. {\bf 411}, 280 (2001).

\bibitem{smerzi97} A. Smerzi, S. Fantoni, S. Giovanazzi, and S. R. Shenoy,
Phys. Rev. Lett. {\bf 79}, 4950 (1997).

\bibitem{milburn97} G. J. Milburn, J. Corney, E. M. Wright, and D. F. Walls,
Phys. Rev. A {\bf 55}, 4318 (1997).

\bibitem{raghavan97} S. Raghavan, A. Smerzi,
S. Fantoni, and S. R. Shenoy, Phys. Rev. A {\bf 59}, 620 (1999).

\bibitem{albiez05} M. Albiez, R. Gati, J. F\"{o}lling, S. Hunsmann, M. Cristiani, and
M. K. Oberthaler, Phys. Rev. Lett. {\bf 95}, 010402 (2005).


\bibitem{wu00} B. Wu and Q. Niu, Phys. Rev. A {\bf 61}, 023402 (2000).

\bibitem{liu02} J. Liu, L. B. Fu, B. Y. Ou, S. G. Chen,
D. I. Choi, B. Wu, and Q. Niu, Phys. Rev. A {\bf 66}, 023404 (2002).

\bibitem{liu03} J. Liu, B. Wu, and Q. Niu, Phys. Rev. Lett. {\bf 90}, 170404 (2003).

\bibitem{theocharis06} G. Theocharis, P. G. Kevrekidis, D. J. Frantzeskakis, and
P. Schmelcher, Phys. Rev. E {\bf 74}, 056608 (2006).

\bibitem{graefe06} E. M. Graefe, H. J. Korsch, and D. Witthaut, Phys. Rev. A {\bf 73},
013617 (2006).

\bibitem{wang06} G. F. Wang, D. F. Ye, L. B. Fu, X. Z. Chen, and J. Liu, Phys.
Rev. A {\bf 74}, 033414 (2006).

\bibitem{itin07} A. P. Itin and S. Watanabe, Phys. Rev. E {\bf 76}, 026218 (2007).

\bibitem{altland09} A. Altland, V. Gurarie, T. Kriecherbauer, and A. Polkovnikov,
Phys. Rev. A {\bf 79}, 042703 (2009).

\bibitem{smith09} K. Smith-Mannschott, M. Chuchem,
M. Hiller, T. Kottos, and D. Cohen, Phys.Rev.Lett. {\bf 102}, 230401 (2009).

\bibitem{ye08} D. F. Ye, L. B. Fu, and J. Liu, Phys. Rev. A {\bf 77}, 013402 (2008).

\bibitem{fu09} L. B. Fu, D. F. Ye, C. H. Lee,
W. P. Zhang, and J. Liu, Phys. Rev. A {\bf 80}, 013619 (2009).

\bibitem{bergmann98} K. Bergmann, H. Theuer, and B.W. Shore,
Rev. Mod. Phys. {\bf 70},1003 (1998).

\bibitem{eckert04} K. Eckert, M. Lewenstein, R. Corbal\'{a}n, G. Birkl, W. Ertmer,
and J. Mompart, Phys. Rev. A {\bf 70}, 023606 (2004).

\bibitem{vitanov06} N. V. Vitanov and B. W. Shore,
Phys. Rev. A {\bf 73}, 053402 (2006).

\bibitem{rab08} M. Rab, J. H. Cole, N. G. Parker, A. D. Greentree, L. C. L.
Hollenberg, and A. M. Martin, Phys. Rev. A {\bf 77}, 061602(R) (2008).

\bibitem{nesterenko09} V. O. Nesterenko, A. N. Nikonov,
F. F. de Souza Cruz, and E. L.
Lapolli, Laser Phys. {\bf 19}, 616 (2009).

\bibitem{ottaviani10} C. Ottaviani, V. Ahufinger, R. Corbalan,
and J. Mompart, Phys. Rev. A {\bf 81}, 043621 (2010).

\bibitem{zhang08} Qi Zhang, P. H\"{a}nggi, and Jianbin Gong,
Phys. Rev. A {\bf 77}, 053607(2008); Jiangbin Gong, Luis
Morales-Molina, and Peter H\"{a}nggi,  Phys. Rev. Lett. {\bf 103},
133002 (2009); L. Morales-Molina and J.B. Gong, Phys. Rev. A {\bf
78}, 041403 (2008).

\bibitem{simon83} B. Simon, Phys. Rev. Lett. {\bf 51}, 2167 (1983).

\bibitem{berry84} M. V. Berry, Proc. R. Soc. London A {\bf 392}, 45 (1984).

\bibitem{born28} M. Born and V. Fock, Z. Phys. {\bf 51}, 165 (1928).

\bibitem{strogatz94} S. H. Strogatz,  Nonlinear Dynamics
and Chaos (Addison-Wesley, New York, 1994).

\bibitem{boyce97} W. E. Boyce and R. C. Diprima, Elementary Differential
Equations and Boundary Value Problems (Wiley, New York, 1997).

\end{references}
\end{document}